\documentclass[english,superaddress,twocolumn]{revtex4}
\usepackage{mathptmx}

\usepackage[T1]{fontenc}
\usepackage[latin9]{inputenc}
\usepackage{color}
\usepackage{graphicx}
\usepackage{esint}

\makeatletter

\providecommand{\tabularnewline}{\\}

\@ifundefined{textcolor}{}
{%
 \definecolor{BLACK}{gray}{0}
 \definecolor{WHITE}{gray}{1}
 \definecolor{RED}{rgb}{1,0,0}
 \definecolor{GREEN}{rgb}{0,1,0}
 \definecolor{BLUE}{rgb}{0,0,1}
 \definecolor{CYAN}{cmyk}{1,0,0,0}
 \definecolor{MAGENTA}{cmyk}{0,1,0,0}
 \definecolor{YELLOW}{cmyk}{0,0,1,0}
}

\makeatother

\usepackage{babel}
\begin{document}

\title{Elliptic flow of thermal dileptons in event-by-event hydrodynamic
simulation}

\author{Hao-jie Xu}

\email{haojiexu@ustc.edu.cn}

\affiliation{Department of Modern Physics, University of Science and Technology
of China, Anhui 230026, People's Republic of China}

\author{Longgang Pang}

\email{lgpang@iopp.ccnu.edu.cn}

\affiliation{Institute of Particle Physics and Key Laboratory of Quarks and Lepton
Physics (MOE), Central China Normal University, Wuhan 430079, People's
Republic of China}

\author{Qun Wang}

\email{qunwang@ustc.edu.cn}

\affiliation{Department of Modern Physics, University of Science and Technology
of China, Anhui 230026, People's Republic of China}
\begin{abstract}
The elliptic flows of thermal di-electrons are investigated within
a (2+1)-dimension event-by-event hydrodynamic model for Au+Au collisions
at $\sqrt{s_{NN}}=200$ GeV. The fluctuating initial conditions are
given by the Monte Carlo Glauber model. We find in our event-by-event
calculation rather weak correlation between the dilepton emission
angles and the event plane angles of charged hadrons. We observe strong
fluctuation effects in dilepton elliptic flows when using the event
plane angles of dileptons at specific invariant mass $M$. The correlation
between the dilepton event plane angle and charged hadron one becomes
stronger with decreasing $M$. This provides a possible measure of the interplay
between the effect of geometric deformation and that of fluctuating
``hot spots'' in relativistic heavy ion collisions.
\end{abstract}


\maketitle

\section{Introduction}

One of the most exciting phenomena found in high-energy heavy ion
collisions (HIC) at Relativistic Heavy Ion Collider (RHIC) and the
Large Hadron Collider (LHC) is the strong collective flow of final
charged hadrons \cite{Ollitrault:1992bk}. This is developed by fast
expansion of Quark Gluon Plasma (QGP) and the Hadron Resonance Gas
(HRG) in the early/later stage of the HIC. The collective flow not
only contributes to the transverse momentum spectra of hadrons but
also leads to the anisotropic distribution in the transverse plane.
This anisotropy is the consequence of the pressure gradient in non-central
collisions which drives the initial geometric asymmetry of the overlapping
region of the colliding nuclei to the momentum anisotropy of final
hadrons. The second Fourier coefficient of the azimuthal distribution
is called the elliptic flow. The big elliptic flow measured at RHIC
and LHC suggests fast local thermalization and the formation of strongly
coupled QGP close to ideal fluid \cite{Ackermann:2000tr,Adcox:2002ms,Aamodt:2010pa}.
So the space-time evolution of the hot and dense matter can be described
by hydrodynamic equations \cite{Kolb:2000sd}.

One critical input in hydrodynamic simulation is the initial condition.
Recently many models such as Monte Carlo Glauber, Monte Carlo Color
Glass Condensate, UrQMD, AMPT and EPOS are used to produce fluctuating
initial conditions for event-by-event (EBE) hydrodynamic simulations
\cite{Hirano:2009ah,Steinheimer:2009nn,Werner:2010aa,Pang:2012he}.
These EBE calculations can explain odd harmonic flows like the triangle
flow and the ridge structure in two-dimensional di-hadron correlation\cite{Pang:2013pma},
they can also provide much better fit to the transverse momentum spectra
and elliptic flows of charged hadrons at RHIC and LHC energy in both
central and semi-central collisions \cite{Alver:2008zza,Holopainen:2010gz,Jia:2012sa,Qiu:2011iv,Qiu:2012uy,Schenke:2010rr}
than smooth one-shot (SM) hydrodynamic simulation with smooth initial
conditions.

Different from the strongly interacting hadrons which escape from
the fireball only on the equal-temperature hyper-surface at freeze-out,
the electromagnetic probes like photons and dileptons, due to their
instant emission once produced, are expected to provide undistorted
information about the space-time information of the QGP and HRG matter
\cite{Chatterjee:2011dw,Chatterjee:2013naa,Kajantie:1986dh,Linnyk:2011vx,McLerran:1984ay,Ruppert:2007cr,Shen:2013cca,Xu:2011tz,vanHees:2007th,vanHees:2006ng,Vujanovic:2013jpa,Xu:2013uza,Adamczyk:2014lpa}.
The invariant mass spectra of dileptons are usually divided into the
low, intermediate and high mass regions (LMR, IMR and HMR) based on
the notion that each region is dominated by different sources of dileptons,
thus provide a method to identify the different evolution stages of
the expanding fireball.

The elliptic flows of dileptons have been studied in SM hydrodynamic
simulations with smooth initial conditions\cite{Mohanty:2011fp,Deng:2010pq,Chatterjee:2007xk}.
However, the production rates of dileptons are sensitive to high temperature
``hot spots'' (fluctuations) in hot and dense medium. So the fluctuations
are expected to play an important role in elliptic flows of dileptons
in heavy ion collisions, which, to our knowledge, have never been
investigated before. The aim of this paper is to calculate the elliptic
flows of dileptons in an EBE hydrodynamic simulation with fluctuating
initial conditions.

This paper is organized as follows. In Sec.\ref{sec:EBE}, we describe
our EBE hydrodynamic model and use it to reproduce the transverse
momentum spectra and elliptic flows of charged hadrons and compare
to RHIC data. In Sec.\ref{sec:Dilepton} we calculate the invariant
mass spectra and elliptic flows of dileptons from the EBE hydrodynamic
simulation. We show the effect of fluctuations on dilepton elliptic
flows with respect to the event plane defined by invariant mass dependent
dilepton spectra and that defined by charged hadron spectra. We give
a summary of our results in the final section.

\section{Event-by-event hydrodynamic model}

\label{sec:EBE}

We use a (2+1)-dimension ideal hydrodynamic model for the EBE simulation.
Our model is similar to the AZHYDRO model \cite{Kolb:2000sd,Pang:2012he}
which implements FCT-SHASTA algorithm\cite{boris1973flux,zalesak1979fully},
but our codes are written in C++. The equation of state (EOS) with
first order phase transition in the original AZHYDRO is replaced by
Lattice QCD EOS which is parameterized as S95P-PCE-v0 \cite{Huovinen:2009yb}
with the chemical freeze-out temperature $T_{chem}=165$ MeV. When
the temperature of the medium is smaller than the kinetic freeze-out
temperature $T_{f}$ (which is set to $120$ MeV in this study), the
spectra of the directly produced hadrons are calculated by the Cooper-Frye
formula \cite{Cooper:1974mv} that mesons/baryons are emitted from
the freeze-out hyper-surface defined by $T_{f}$ following the Bose-Einstein/Fermi-Dirac
distribution. We take more resonances \cite{Eidelman:2004wy} into
account in our model than AZHYDRO. In order to calculate the dilepton
spectra, we keep track of the local energy density and flow velocity
profiles for the QGP and HRG matter during the evolution.

The Monte Carlo Glauber model \cite{Hirano:2009ah} is employed to
generate the profile of the fluctuating initial entropy density in
Au+Au collisions at $\sqrt{s_{NN}}=200$ GeV. In the Glauber model,
the initial entropy density is assumed to be proportional to a linear
combination of the number of participants $dN_{\mathrm{part}}/d^{2}\mathbf{x}_{\perp}$and
that of binary collisions $dN_{\mathrm{coll}}/d^{2}\mathbf{x}_{\perp}$\cite{Hirano:2009ah},
\begin{equation}
\left.\frac{ds}{\tau_{0}dxdyd\eta_{s}}\right|_{\eta_{s}=0}=\frac{C}{\tau_{0}}\left(\frac{1-\delta}{2}\frac{dN_{\mathrm{part}}}{d^{2}\mathbf{x}_{\perp}}+\mbox{\ensuremath{\delta}}\frac{dN_{\mathrm{coll}}}{d^{2}\mathbf{x}_{\perp}}\right),
\end{equation}
where $C=16$ and $\delta=0.14$ are two constants, and the initial
time is chosen to be $\tau_{0}=$0.4 fm/c. Since we use (2+1)-dimension
hydrodynamic model with Bjorken boost-invariance \cite{Bjorken:1982qr}
and focus on the central rapidity region, it is convenient to set
the spatial rapidity $\eta_{s}$ zero.

Normally the freeze-out hyper-surfaces are calculated in cuboidal way
\cite{Kolb:2000sd,Hirano:2001eu,Schenke:2010nt,Pang:2012he} for each
fluid cell. In the AZHYDRO model, in order to save the computing time,
several fluid cells are used to generate a big piece of freeze-out
hyper-surface. The flow velocity at the hyper-surface obtained in this
way is an average of these cells. This will introduce numerical errors
when the flow velocities are different for neighboring cells (especially
with fluctuating initial conditions). In our model, we used a better
treatment of the freeze-out hyper-surface. We first discretize the
fluid velocity $u^{\sigma}$ and then take the sum over the hyper-surface
elements $\sum_{u_{i}^{\sigma}\in[u^{\sigma},u^{\sigma}+du^{\sigma}]}d\Sigma_{\mu}^{i}$
for each fluid cell whose velocity is inside the bin $[u^{\sigma},u^{\sigma}+du^{\sigma}]$
before carrying out the sum over different $u^{\sigma}$ in the Cooper-Frye
formula. This will substantially reduce the computing time while keeping
the numerical accuracy with appropriately chosen bin size of the fluid
velocity (the bin size is set to 0.02 in this paper). This method
has been checked that it can reproduce the transverse momentum spectra
and elliptic flows of hadrons.

The harmonic coefficients $v_{n}$ which depend on transverse momenta
are defined by
\begin{equation}
v_{n}(p_{T})=\frac{\int_{0}^{2\pi}d\phi\frac{dN}{p_{T}dp_{T}d\phi dy}\cos[n(\phi-\Psi_{n})]}{\int_{0}^{2\pi}d\phi\frac{dN}{p_{T}dp_{T}d\phi dy}},
\end{equation}
where $\Psi_{n}$ are azimuthal angles of the event plane in momentum
space
\begin{equation}
\Psi_{n}=\frac{1}{n}\arctan\frac{\langle p_{T}\sin(n\phi)\rangle}{\langle p_{T}\cos(n\phi)\rangle}.\label{eq:EP}
\end{equation}
Here the average is taken over final charged hadrons in one event.
The second and third harmonic coefficient are often called elliptic
and triangle flow.

\begin{figure}
\begin{centering}
\includegraphics[scale=0.45]{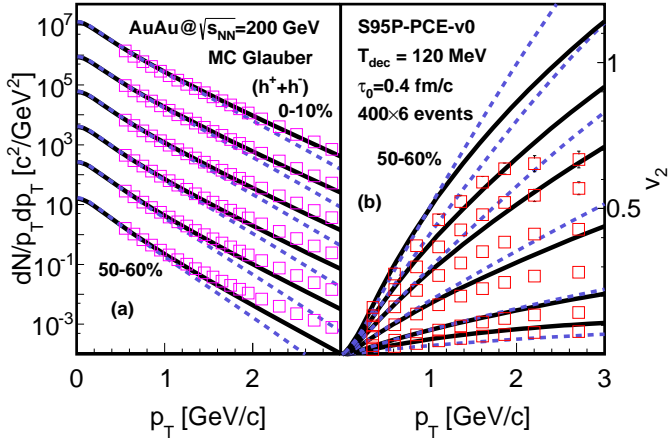}
\par\end{centering}

\caption{(Color online) (a) Transverse momentum spectra and (b) elliptic flows
$v_{2}$ for charged hadrons as functions of $p_{T}$ in different
centrality bins for Au+Au collisions at $\sqrt{s_{NN}}=200$ GeV.
The hydrodynamic simulations are done with fluctuating (black-solid
lines) and smooth(blue-dash lines) initial conditions. In the top-down
order, the curves correspond to centrality bins (a) $(0-10\%)\times10^{4}$,
$(10-20\%)\times10^{3}$, $(20-30\%)\times10^{2}$, $(30-40\%)\times10$,
$(40-50\%$), $(50-60\%)\times0.1$ and (b) $(50-60\%)\times3$, $(40-50\%)\times2.5$,
$(30-40\%)\times2$, $(20-30\%)\times1.5$, $(10-20\%)$, $(0-10\%)$,
where the number outside the parenthesis denotes the normalization
factor of each curve. The data (symbols) are taken from the PHENIX
collaboration\cite{Adler:2003au,Adare:2011tg}. \label{fig:hadron}}

\end{figure}

In Fig. \ref{fig:hadron} we show the $p_{T}$ spectra and elliptic
flow of charged hadrons for both EBE (400 events for each centrality bins) and SM hydrodynamic simulation
in a variety of centralities. The centrality bins are determined by
the number of participants $N_{\mathrm{part}}$ in Tab.\ref{tab:Centralitytable}.
The results are compared to the PHENIX data \cite{Adler:2003au,Adare:2011tg}.
In the SM case, the event-averaged initial entropy density distributions
are used in the hydrodynamic calculation. We use the reaction plane
as reference to take event average in the SM simulation. Another method
is to use the participant plane. The two methods are different in
elliptic flow for most central collisions \cite{Qiu:2011iv,Schenke:2010rr,Pang:2012he}.

\begin{table}
\caption{\label{tab:Centralitytable}Centralities and numbers of participants
for Au+Au@$\sqrt{s}=200$ GeV calculated in one million events. The
centrality bins are determined by number of participants $N_{\mathrm{part}}$.
Here $N_{\mathrm{part}}^{\mathrm{min}}$ corresponds to the upper
bound of the centrality bin. }

\begin{centering}
\begin{tabular}{|c|c|c|c|c|c|c|}
\hline
Cnt & 0-10\% & 10-20\% & 20-30\% & 30-40\% & 40-50\% & 50-60\%\tabularnewline
\hline
\hline
$N_{\mathrm{part}}^{\mathrm{min}}$ & 277 & 199 & 140 & 95 & 61 & 37\tabularnewline
\hline
\end{tabular}
\par\end{centering}

\end{table}

We see in Fig. \ref{fig:hadron}(a) that the $p_{T}$ spectra of charged
hadrons in the EBE simulation is harder than the SM simulation in
each centrality bin. The better agreement of the EBE result with data
is due to the bigger pressure gradient produced from the ``hot spots''
which drives bigger collective flows in the early stage of hydrodynamic
evolution \cite{Qiu:2011iv,Pang:2012he}. Charged hadron spectra are
only sensitive to those ``hot spots'' which are close to the freeze-out
hyper-surface, while thermal photon and dilepton spectra are influenced
by all ``hot spots'' in whole space-time volume.

In Fig. \ref{fig:hadron}(b), we see that the elliptic flows in the
EBE case are lower than the SM results in semi-central and peripheral
collisions, while they are higher than the SM results in central collisions.
This clearly shows the fluctuation effect on the elliptic flow. We
know that the elliptic flow is the result of geometric eccentricity
and initial fluctuation. The geometric eccentricity gives a non-zero
elliptic flow, while hot spot fluctuation gives almost an isotopic
flow. In semi-central and peripheral collisions, the geometric eccentricity
in the SM case is slightly smaller than that in the EBE case. The
fluctuation effect greatly reduces the elliptic flow in the EBE case.
In central collisions, the deviation of the participant plane from
the reaction plane is larger for central collisions than peripheral
collisions, so the eccentricity in the SM case is much smaller than
that in the EBE case \cite{Schenke:2010rr}, which gives much smaller
elliptic flow in the SM case even with the reduction effect from fluctuation
in the EBE calculation.

\begin{figure}
\begin{centering}
\includegraphics[scale=0.45]{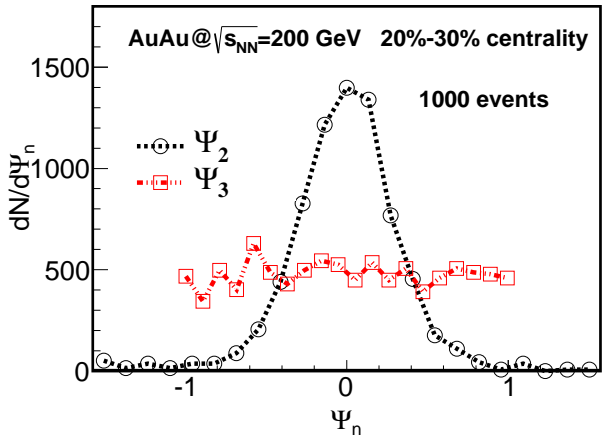}
\par\end{centering}

\caption{(Color online) The 2nd and 3rd event plane azimuthal angle distribution
from EBE hydrodynamic simulation with 1000 fluctuating events.\label{fig:eventplane} }
\end{figure}

To identify whether the harmonic flow is generated by geometric eccentricity
or fluctuating ``hot spots'', it is worth to study the azimuthal
angle distribution of the event plane. In our hydrodynamic simulation,
the reaction plane's azimuthal angle is set to 0. In Fig. \ref{fig:eventplane},
we show the distribution associated with the second and third harmonic
flows at centrality $20-30\%$ by using 1000 hydrodynamic events with
fluctuating initial energy density distributions. We see that $\Psi_{3}$
are completely uncorrelated with the reaction plane. The de-correlation
indicates that the triangle flows are mostly generated by fluctuating
``hot spots''. However $\Psi_{2}$ are strongly correlated with
the reaction plane as expected since $v_{2}$ is actually a collective
flow response to the geometric deformation \cite{Qiu:2011iv} in semi-central
and peripheral collisions.

\section{Elliptic flow of thermal dileptons }

\label{sec:Dilepton}

The differential production rate of dileptons per unit volume can
be written in the following form (see, e.g. \cite{McLerran:1984ay,vanHees:2007th}),
\begin{eqnarray}
\frac{dN}{d^{4}xd^{4}p} & = & -\frac{\alpha}{4\pi^{2}}\frac{1}{M^{2}}n_{B}(p\cdot u)\left(1+\frac{2m_{l}^{2}}{M^{2}}\right)\nonumber \\
 &  & \times\sqrt{1-\frac{4m_{l}^{2}}{M^{2}}}\mathrm{Im}\Pi^{R}(p,T).
\end{eqnarray}
Here $m_{l}$ is the lepton mass, $\alpha$ is fine structure constant,
$p$ is the dilepton's four momentum and $M\equiv\sqrt{p^{2}}$ is
the dilepton invariant mass, $n_{B}(p\cdot u)\equiv1/(\exp(p\cdot u/T)-1)$
($T$ and $u$ are the local temperature and fluid velocity, respectively)
is the Bose-Einstein distribution function, and $\Pi^{R}$ is the
retarded polarization tensor from the quark loop in the QGP phase
or the hadronic loop in the HRG phase. In the HRG phase, the thermal
dilepton production rate is dominated by in-medium $\rho$ meson decays.\emph{
}The details of the calculation of dilepton invariant mass spectra
in Au+Au collisions at the RHIC energy are given in Ref. \cite{Deng:2010pq,Xu:2011tz}
by some of us. To simplify our calculation, we will focus on $20-30\%$
centrality bin and neglect the contribution from in-medium $\omega$
and $\phi$ meson decays.

\begin{figure}
\begin{centering}
\includegraphics[scale=0.45]{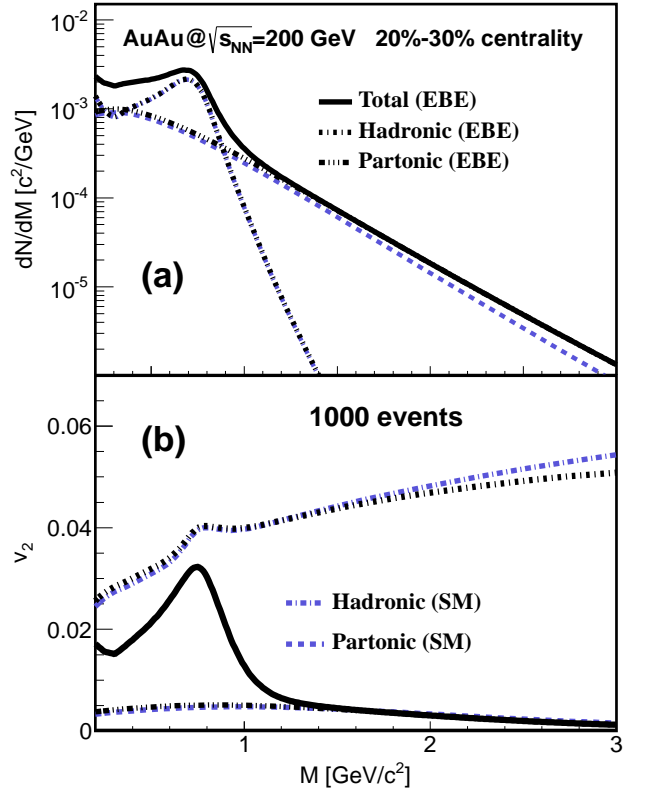}
\par\end{centering}

\caption{(Color online) (a) Invariant mass spectra and (b) elliptic flows of
di-electrons in semi-central Au+Au collisions at  $\sqrt{s_{NN}}=$200
GeV.\label{fig:dilepton}}
\end{figure}

In Fig. \ref{fig:dilepton}(a) we show the invariant mass spectra
of di-electrons in semi-central ($20-30\%$ centrality bin) Au+Au
collisions at the RHIC energy. The LMR thermal dilepton production
rate is dominated by in-medium $\rho$ decays. The broadened dilepton
invariant mass spectra around the mass of the free $\rho$ indicate
a strong medium modification by scatterings between $\rho$ and other
mesons and baryons in thermal medium. This modification is described
by using hadronic many body theory\cite{Rapp:2000pe} and empirical
scattering amplitude method \cite{Eletsky:2001bb}\emph{.} In the
IMR the dileptons from QGP are dominant, and more dileptons are produced
in the QGP phase in the EBE calculation (dash-dot-dot-dotted line,
overlapping with the solid line) than the SM calculation (dashed).
Such an enhancement is due to the ``hot spots'' in fluctuating initial
conditions where larger-than-average temperature give a large contribution
to the HMR dieltpon. Similar results were found in the EBE calculation
for thermal photons \cite{Chatterjee:2011dw}. The fluctuation effect
on the invariant mass spectra in the HRG phase is negligible, since
most of hadronic dileptons are produced later than the partonic ones
\cite{Deng:2010pq} and the HRG phase only appears below the transition
temperature.

The differential elliptic flow coefficient $v_{2}(M)$ as a function
of the invariant mass is given by
\begin{equation}
v_{2}(M)=\frac{\int d\phi(dN/dMd\phi dy)\cos(2(\phi-\Psi_{2}))}{\int d\phi(dN/dMd\phi dy)},\label{eq:dileptonv2}
\end{equation}
where $\Psi_{n}$ with $n=2$ is the azimuthal angle of the event
plane for final state charged hadrons in momentum space as defined
in Eq. (\ref{eq:EP}). The $v_{2}(M)$ results are shown in Fig. \ref{fig:dilepton}(b).
We see in the figure that the elliptic flows of dileptons in the HRG
phase in the EBE and SM case increase with $M$. The small peak in
the hadronic component around the $\rho$ mass is due to the temperature-dependent
in-medium $\rho$ meson spectral functions. The elliptic flow in the
QGP phase is much smaller than that in the HRG phase, since the QGP
phase is in the early stage of the fireball evolution before the transverse
flow is fully developed. The sharp decrease of the elliptic flow from
the HGR dominated region to the QGP dominated region can be a possible
signal for QGP formation in heavy ion collisions \cite{Deng:2010pq}.

\begin{figure}
\begin{centering}
\includegraphics[scale=0.45]{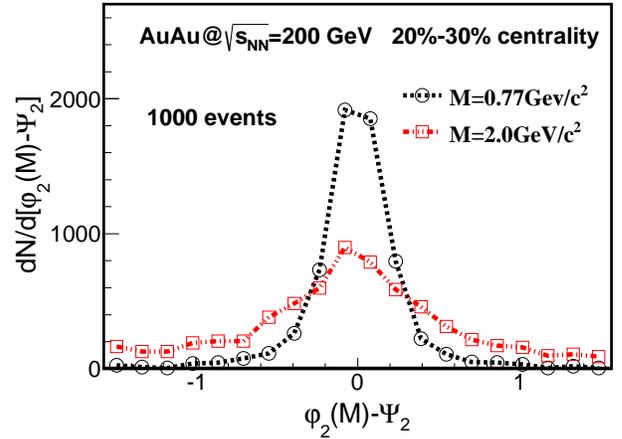}
\par\end{centering}

\caption{\label{fig:dileptonEP}(Color online) The distribution of $\varphi_{2}-\Psi_{2}$
at dilepton invariant mass $M=0.77$ GeV/$c^{2}$ and $M=2$ GeV/$c^{2}$
in EBE hydrodynamic simulation with the same 1000 events as in Fig.
(\ref{fig:eventplane}). $\varphi_{2}$/$\Psi_{2}$ is the event plane
azimuthal angle defined by dileptons/hadrons. }
\end{figure}

\textcolor{black}{As shown in Fig. \ref{fig:dilepton}(b), the elliptic
flow with the fluctuating initial condition in the LMR/HMR is slightly
larger/smaller than that without fluctuation. Such a small fluctuation
effect is beyond our expectation since most IMR and HMR dileptons are
produced at early time of the fireball expansion and their harmonic
flow should be sensitive to the initial fluctuating ``hot-spots''.
The reason is that }the dilepton elliptic flow is calculated with
respect to the event plane azimuthal angle $\Psi_{2}$ for final charged
hadrons according to Eq. (\ref{eq:dileptonv2}), however $\Psi_{2}$
is strongly correlated to the initial geometric deformation but not
to dilepton azimuthal angles with the fluctuating initial condition.

To support the above reasoning, we introduce the following event plane
azimuthal angle $\varphi_{n}(M)$ for dileptons,
\begin{equation}
\varphi_{n}(M)=\frac{1}{n}\arctan\frac{\langle p_{T}\sin(n\phi)\rangle}{\langle p_{T}\cos(n\phi)\rangle}.
\label{eq:EPdilepton}
\end{equation}
Here the average is taken over thermal dileptons at specific mass in one
event. In Fig. (\ref{fig:dileptonEP}) we show the distribution of $\varphi_{2}(M)-\Psi_{2}$
for thermal dileptons at $M=0.77$ GeV/$c^{2}$ (hadronic dominated
region) and $M=2.0$ GeV/$c^{2}$ (partonic dominated region) with
the same events as in Fig. (\ref{fig:eventplane}). It is obvious
that the correlation of $\varphi_{2}(M)$ and $\Psi_{2}$ is much
stronger in the HRG phase than that in the QGP phase. In this case,
when we choose the event plane angle $\Psi_{2}$ of charged hadrons
to calculate the dilepton elliptic flow at $M=2.0$ GeV/$c^{2}$,
the fluctuation effect is washed out by the weak correlation of $\varphi_{2}(M)$
and $\Psi_{2}$. The de-correlation of $\varphi_{2}(M)$ and $\Psi_{2}$
is due to the interplay of the geometric deformation and fluctuating
initial ``hot spots''. In the early stage, the elliptic flow is
small and the de-correlation effect is large, while in the later hadronic
stage, the elliptic flow is driven by the geometric deformation and
the de-correlation effect becomes weaker. The correlation between $\varphi_{2}(M)$
and $\Psi_{2}$ becomes stronger from higher mass to lower mass. This
provides us with a measure of the interplay between the effect of
geometric deformation and that of fluctuating ``hot spots'' in relativistic
heavy ion collisions.

\begin{figure}
\begin{centering}
\includegraphics[scale=0.45]{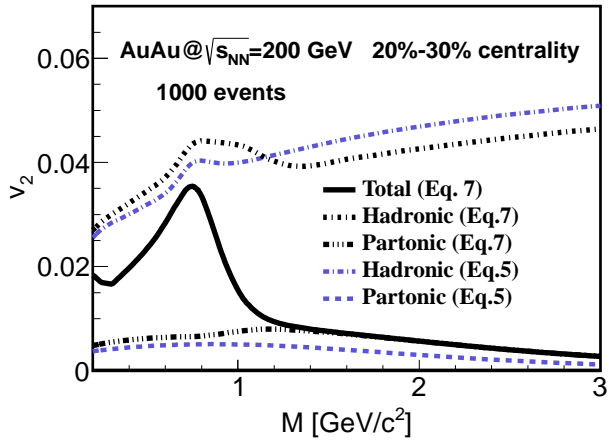}
\par\end{centering}

\caption{(Color online) The elliptic flows of di-electrons in semi-central
Au+Au collisions at $\sqrt{s_{NN}}=$200 GeV with the event plane
azimuthal angles defined by dileptons at specific $M$. \label{fig:v2New}}

\end{figure}

To look at the effect of replacing $\Psi_{2}$ with $\varphi_{2}(M)$,
we can define the dilepton elliptic flow as
\begin{equation}
v_{2}(M)=\frac{\int d\phi(dN/dMd\phi dy)\cos(2(\phi-\varphi_{2}(M))}{\int d\phi(dN/dMd\phi dy)}.\label{eq:v2New}
\end{equation}
Fig. (\ref{fig:v2New}) shows the big difference between the $v_{2}(M)$
results with $\varphi_{2}(M)$ and that with $\Psi_{2}$ in the EBE
hydrodynamic simulation. This difference is due to the strong de-correlation
between $\Psi_{2}$ and $\varphi_{2}(M)$. Notice that the contribution from hadronic phase to Eq. (\ref{eq:v2New}) is below that to Eq. (\ref{eq:dileptonv2}) at $M>1.2$ GeV/$c^2$. This is because the dilepton production rate at high mass region is dominated by the partonic contribution, the dilepton event plane defined by Eq. (\ref{eq:EPdilepton}) in that mass region is almost that of dileptons from partonic phase instead of hadronic phase. Therefore when looking at $v_{2}$ of dileptons from hadronic phase in high mass region, it is more correlated to the orientation of charged hadrons than to that of dileptons mainly from partonic phase. Mathematically the mean  value of $|\varphi_{h}(M) - \varphi_2(M)|$ in Eq. (\ref{eq:v2New}) is bigger than that of $|\varphi_{h}(M) - \Psi_2|$ in Eq. (\ref{eq:dileptonv2}) at high mass region which will bring a smaller $v_2$, where $\varphi_{h}(M)$ is the azimuthal angle of dilepton from hadronic phase at specific mass. The de-correlation suggests
that the importance of the choices of event planes in maximizing the
dilepton elliptic flows. In experiments, however, it is difficult
to identify the dilepton event planes because of very low production
rates dileptons.

The de-correlation effects are also present for final charged hadrons.
In a recent work on two-particle correlation \cite{Gardim:2012im},
it was pointed out that the flow angle $\Psi_{n}$ may depend on the
transverse momentum $p_{T}$ and the pseudo-rapidity $\eta$. The $p_{T}$-dependent
event plane angle for charged hadrons have also been studied in Ref.
\cite{Heinz:2013bua}. Similar de-correlation effect was found in thermal
photon elliptic flows with viscous hydrodynamic simulation \cite{Shen:2013cca},\emph{
}where the event angle $\Psi_{n}$ from pions de-correlate from $\Psi_{n}^{\gamma}(p_{T})$
defined by thermal photons at specific $p_{T}$. While in our work,
we realize the evolution of de-correlation effect in relativistic heavy
ion collisions.

\section{Summary and conclusion}

\label{sec:summary}

We investigated the elliptic flows of di-electrons in Au+Au collisions
at $\sqrt{s_{NN}}=200$ GeV by using a (2+1)-dimension event-by-event
hydrodynamic model. The fluctuating initial entropy density profile
is generated by the Monte Carlo Glauber model. The event-by-event
hydrodynamic simulation gives a better description of transverse momentum
spectra and elliptic flows of final charged hadrons in central and
semi-central collisions. The event plane angle distribution indicates
that the elliptic flow is largely generated by the initial geometric
deformation while the triangle flow is largely generated by initial
fluctuating ``hot spots''.

The dilepton invariant mass spectra are harder in the HMR from event-by-event
hydrodynamic simulation than the one-shot results, which is due to
larger-than-average temperatures of fluctuating ``hot spots''. The
fluctuation effects are small when we use the event plane angle from
final charged hadrons in calculating dilepton elliptic flows.

We observed bigger fluctuation effects when we used event plane angles
defined by dileptons at specific $M$. The correlation between the
event plane angle of dileptons at specific $M$ and that of charged
hadrons becomes stronger with decreasing $M$. This provides us with
a possible measure of the interplay between the effect of geometric
deformation and that of fluctuating ``hot spots'' in relativistic
heavy ion collisions.
\begin{acknowledgments}
This work is supported partially by the National Natural Science Foundation
of China projects (No. 11125524) and the Major State Basic Research
Development Program in China (No. 2014CB845402 )
\end{acknowledgments}
\bibliographystyle{apsrev}
\bibliography{Ref}

\end{document}